\documentstyle [12pt]{article}
\begin{document}
\baselineskip 20.0 pt
\par
\mbox{}
\vskip -1.25in\par
\mbox{}
\begin{flushright}
\makebox[1.5in][1]{UU-HEP-92/11}\\
\makebox[1.5in][1]{May 1992}\\
 \end{flushright}
\vskip 0.25in

\begin{center}
{\bf An Infinite Number of Commuting Quantum $\hat{W}_{\infty}$ Charges}\\
{\bf in the $SL(2,R)/U(1)$ Coset Model}\\
\vspace{40 pt}
{Feng Yu and Yong-Shi Wu}\\
\vspace{20 pt}
{{\it Department of Physics, University of Utah}}\\
{{\it Salt Lake City, Utah 84112, U.S.A.}}\\
\vspace{40 pt}
{\bf Abstract}\\
\end{center}
\vspace{10 pt}

The conformal non-compact $SL(2,R)/U(1)$ coset model in two dimensions
has been recently shown to embody a nonlinear $\hat{W}_\infty$ current
algebra, consisting of currents of spin $\geq 2$ including the
energy-momentum tensor. In this letter we explicitly construct an
infinite set of commuting quantum $\hat{W}_\infty$ charges in the model with
$k=1$. These commuting quantum charges generate a set of infinitely many
compatible flows (quantum KP flows), which maintain the nonlinear
$\hat{W}_\infty$ current algebra invariant.

\newpage

In the last a few years the exploitation of infinite dimensional
algebras, such as Virasoro algebra [1], Kac-Moody current algebras [2]
and extended conformal $W_N$ algebras [3], has played a crucial role
in solving several large classes of two-dimensional quantum
conformal field theories,
e.g. the minimal models [4], Wess-Zumino-Witten models [5] and compact coset
models [6]. Very recently a new nonlinear current algebra, called the
$\hat{W}_\infty$ algebra [7], is shown to hide in the 2d non-compact
$SL(2,R)/U(1)$ coset model both at the classical [8] and the
quantum [9,8] levels. This algebra is generated by currents of all spin
$s \geq 2$, including the energy-momentum tensor, and
can be viewed as a generalization of Zamolochikov's nonlinear $W_N$
algebra [3] which is generated by currents of spin $s$ from 2 to $N$. At the
classical level, the $\hat{W}_\infty$ algebra is shown [8] to be a
nonlinear, centerless deformation of the linear $W_\infty$
algebra [10,11], and isomorphic to the second Hamiltonian
structure of the KP hierarchy [12] proposed by Dickey [13]. This
isomorphism immediately suggests the existence of infinitely many
charges that are in involution in the classical
$SL(2,R)/U(1)$ coset model [8].

However, after quantizing the model the $\hat{W}_\infty$
current algebra, like the $W_N$ algebras, undergoes nontrivial
quantum deformation and receives intriguing quantum corrections:
The current-current commutators acquire not only non-vanishing central
terms, but also additional linear and nonlinear terms which have
non-trivial dependence on the level $k$ of the coset
model.  Despite these complications, it has been conjectured in ref. [9,8]
that there exists an {\it infinite} set of {\it commuting quantum}
$\hat{W}_\infty$ charges in the quantized model. This letter is devoted
to showing the validity of this conjecture, by explicit construction of
the commuting quantum charges in the model with level $k=1$.

Let us start with a brief review for the classical $\hat{W}_{\infty}$
current algebra [8] in the $SL(2,R)_{k}/U(1)$ model. Essentially this is a
two-boson realization of the $\hat{W}_\infty$ algebra. Consider
only the holomorphic part. The $SL(2,R)_k$ current algebra is known
[14] to have a three-boson realization as follows:
\begin{eqnarray}
J_{\pm}=\sqrt{\frac{k}{2}} e^{\pm \sqrt{\frac{2}{k}} \phi_3}
(\phi_1^{'} \mp i\phi_2^{'}) e^{\pm\sqrt{\frac{2}{k}}\phi_1},~~~
J_3 = - \sqrt{\frac{k}{2}} \phi_3^{'}.
\end{eqnarray}
To take the coset, one simply restricts the $U(1)$
current $J_3=0$ or the boson $\phi_3=0$. Thus one is left with $J_{\pm}$
which are nothing but the bosonized para-fermion currents:
\begin{eqnarray}
J_{+}=\bar{j}e^{\bar{\phi}+\phi}, ~~~~
J_{-}=je^{-\bar{\phi}-\phi}.
\end{eqnarray}
Here $\phi=(1/{\sqrt 2})(\phi_1+i\phi_2)$ and $\bar{\phi}=(1/\sqrt{2})
(\phi_1-i\phi_2)$, and their currents $\bar{j}(z)=\bar{\phi}'(z)$,
$j(z)=\phi'(z)$ satisfy the Poisson brackets
\begin{eqnarray}
& & {\{\bar{j}(z),\bar{j}(w)\}} ={\{j(z), j(w)\}} = 0, \nonumber\\
& & {\{j(z), \bar{j}(w)\}} = \partial_{z}\delta(z-w).
\end{eqnarray}
At the classical level, we can set the level $k$ equal to 1 without
loss of generality.
The classical $\hat{W}_{\infty}$ currents, denoted as $u_r(z)$
($r=0,1,2,\cdots$) with spin $s=r+2$, are generated by the following
(bi-local) product expansion of the coset currents
\begin{eqnarray}
J_{+}(z)J_{-}(z') =
\sum^{\infty}_{r=0}u_{r}(z)\frac{(z-z')^{r}}{r!}.
\end{eqnarray}

It has been shown in ref.[8] that these currents $u_{r}(z)$ are
formed only from the currents $j(z)$, $\bar{j}(z)$ and their derivatives,
containing no fields $\phi$ or $\bar{\phi}$ themselves;
and their Poisson brackets
satisfy the $\hat{W}_{\infty}$ algebra [7]:
\begin{eqnarray}
{\{u_{r}(z), u_{s}(w)\}} = k_{rs}(z)\delta(z-w)
\end{eqnarray}
with $k_{rs}(z)$ explicitly given by
\begin{eqnarray}
k_{rs}^{(2)} &=& \sum^{s+1}_{l=0} \left( \begin{array}{c}
s+1 \\ l
\end{array} \right) D^{l}u_{r+s+1-l} -\sum^{r+1}_{l=0} \left( \begin{array}{c}
r+1 \\ l
\end{array} \right) u_{r+s+1-l}(-D)^{l} \nonumber\\
& & -\sum^{r-1}_{l=0}\sum^{s-1}_{k=0}(-1)^{r-l}\left( \begin{array}{c}
r\\l
\end{array} \right) \left( \begin{array}{c}
s\\k
\end{array} \right) u_{l}D^{r+s-l-k-1}u_{k} \nonumber\\
& & +\sum^{\infty}_{l=0}{[
\sum^{l+s}_{t=l+1}(-1)^{l} \left( \begin{array}{c}
t-s-1 \\ l
\end{array} \right)} \\
& & { - \sum^{r+s}_{t=r+1}\sum^{t-r-1}_{k=0}(-1)^{l+k}
\left( \begin{array}{c}
t-k-1 \\ l-k
\end{array} \right) \left( \begin{array}{c}
s \\ k
\end{array} \right) ]}u_{t-l-1}D^{l}u_{r+s-t}. \nonumber
\end{eqnarray}
As shown in ref.[7], on one hand, this Poisson bracket algebra can be
viewed as a nonlinear deformation of $W_{\infty}$; on the other hand,
it is isomorphic to the second Hamiltonian structure of the KP hierarchy
proposed by Dickey [13].

In the context of the KP hierarchy, it is the coefficient functions
$u_r(z)$ in the KP pseudo-differential operator
\begin{eqnarray}
L = D+\sum^{\infty}_{r=0}u_{r}(z)D^{-r-1},
{}~~~~D\equiv \partial/\partial{z},
\end{eqnarray}
which have the same Poisson brackets (5) and (6). These Poisson
brackets give rise to the second KP Hamiltonian structure, in the
sense that the original KP hierarchy
\begin{eqnarray}
\frac {\partial L}{\partial t_{m}} = {[(L^{m})_{+},L]}
{}~~~(m=1,2,3,\ldots)
\end{eqnarray}
can be put into the Hamiltonian form
\begin{eqnarray}
\frac {\partial u_{r}(z)}{\partial t_{m}} = {\{u_{r}(z),
\oint_{0}H_{m+1}(w)dw\}},
\end{eqnarray}
with the Hamiltonian functions defined by
\begin{eqnarray}
H_{m+1} = \frac{1}{m}ResL^{m},
\end{eqnarray}
where the residue $Res$ means the coefficient of the $D^{-1}$ term.
These KP Hamiltonian functions give rise to an infinite set of conserved
charges in involution [13]:
\begin{eqnarray}
{\{\oint_{0}H_{m}(z)dz, \oint_{0}H_{n}(w)dw\}} =0.
\end{eqnarray}

As shown in ref.[8], the above two-boson realization (4) of the
$\hat{W}_{\infty}$ currents is equivalent to
expressing the KP operator $L$ directly in terms of $\bar{j}$ and $j$
as follows:
\begin{eqnarray}
L = D+\bar{j}\frac{1}{D-(\bar{j}+j)}j
\equiv D+\bar{j}D^{-1}j+\bar{j}D^{-1}(\bar{j}+j)D^{-1}j+ \cdots .
\end{eqnarray}
Since the $\hat{W}_{\infty}$ algebra for the currents $u_{r}$ is
isomorphic to the second KP Hamiltonian structure, by the same
algebraic manipulations as in the KP case we infer that in
the classical $SL(2,R)/U(1)$ model there exists an infinite set of
commuting $\hat{W}_{\infty}$ charges, given by the
integral of the same set of $H_{m+1}$ in eq.(10)
with the $u_r(z)$ in $L$ identified with
those constructed from the generating function (4).

We now proceed to the quantized $SL(2,R)/U(1)$ model and attack the
problem of constructing an infinite set of commuting quantum charges
which are a deformation of the above classical charges. This problem
is a difficult one since in the present case
quantization will introduce quite involved quantum corrections
to the current algebra.

To start we note in the first place that compared to their
classical expression (2), the two boson prescription for the
coset currents $J_{\pm}$ receives additional terms:
\begin{eqnarray}
J_{+}(p,z) &=&
\frac{1}{2}[(1+\sqrt{1-2p})\bar{j}+(1-\sqrt{1-2p})j]
e^{\sqrt{p}(\bar{\phi}+\phi)}, \nonumber\\
J_{-}(p,z) &=&
\frac{1}{2}[(1-\sqrt{1-2p})\bar{j}+(1+\sqrt{1-2p})j]
e^{-\sqrt{p}(\bar{\phi}+\phi)}.
\end{eqnarray}
Here for convenience we use $p=k^{-1}$ as the deformation parameter,
so that the classical limit is $p\rightarrow 0$ after rescaling $\phi
\rightarrow \phi/\sqrt{p}$, $J_\pm \rightarrow J_\pm/\sqrt{p}$ (and $u_r
\rightarrow u_r/p)$. (Recall that in the quantum theory, $p$ actually
represents $p\hbar$ and the field $\phi$ has the dimension $\sqrt{\hbar}$
so does each current.) As usual, the currents $J_{\pm}$
are operators so that the right-hand side of eq.(13) should be
normal-ordered. But from now on we will always suppress the notation
for normal ordering. Furthermore, the quantum $\hat{W}_{\infty}$
currents $u_{r}(z;p)$ are generated by the following operator product
expansion (OPE)
\begin{eqnarray}
J_{+}(p,z)J_{-}(p,z') =
\epsilon^{-2p}\{\epsilon^{-2}+\sum^{\infty}_{r=0}u_{r}(p,z)
\frac{\epsilon^{r}}{r!}\}
\end{eqnarray}
where $\epsilon\equiv z-z'$. This OPE can be viewed as the quantum
deformation of the classical product expansion (4). By substituting
eq.(13) into the left-hand side, it is easy to see that each current
$u_{r}$ on the right-hand side will acquire many new terms, which are
highly nonlinear in $j(z)$ and $\bar{j}(z)$ and
have nontrivial $p$-dependence.

As usual the complete structure of the quantum $\hat{W}_{\infty}(p)$
algebra can be easily read off from the OPE's of the quantum currents
$u_{r}(z;p)u_{s}(w;p)$. For the latter we extract them from the OPE of
the following four $SL(2,R)_{k}$ $/U(1)$ coset currents:
\begin{eqnarray}
(J_{+}(z;p)J_{-}(z-z';p))
(J_{+}(w;p)J_{-}(w-w';p)).
\end{eqnarray}
The first few quantum currents and their OPE's obtained in this way
have been explicitly presented in our previous paper [8]. In contrast to
the classical case, a general expression in the quantized model (with
generic value of $p$) for the commuting Hamiltonians $H_m$ as elegant as
eq.(10) seems beyond our reach. However, very fortunately
we have been able to find a general formula for the commuting quantum charges
in the model with level $k=1$.

When $k$=1 or $p$=1, the computation of the OPE's in eqs.(14) and (15)
becomes rather simplified. The first a few $\hat{W}_{\infty}$ currents
$u_{r}(z)$ read
\begin{eqnarray}
u_{0} &=& -\bar{j}j-\frac{i}{2}(\bar{j}'-j'), \nonumber\\
u_{1} &=& -\frac{1}{2}(\bar{j}^{2}j+\bar{j}j^{2})
+\frac{1}{6}(\bar{j}^{3}+j^{3})
-\frac{i}{2}(\bar{j}\bar{j}'-jj')
-\frac{1}{12}(1-3i)\bar{j}'' \nonumber\\
& & - \frac{1}{12}(1+3i)j'', \nonumber\\
u_{2} &=& -\frac{1}{2}\bar{j}^{2}j^{2} +\frac{1}{4}(\bar{j}^{4}+j^{4})
+\bar{j}j(\bar{j}'+j') -\frac{1}{2}(1+i)(\bar{j}^{2}-j^{2})\bar{j}'
\nonumber\\
& & +\frac{1}{2}(1-i)(\bar{j}^{2}-j^{2})j' -\frac{1}{2}\bar{j}j''
-\frac{1}{2}\bar{j}''j -\frac{1}{2}\bar{j}'j' -\frac{1}{4}(1-2i)\bar{j}'^{2}
\nonumber\\
& & -\frac{1}{4}(1+2i)j'^{2} +\frac{i}{2}(\bar{j}\bar{j}''-jj'')
+\frac{1}{12}(1-2i)\bar{j}''' +\frac{1}{12}(1+2i)j'''.
\end{eqnarray}
What makes the $p=1$ case
manipulable is that without too much difficulty from the OPE (15)
we can extract the crucial $(z-w)^{-1}$ terms in the following OPE's:
\begin{eqnarray}
u_{0}(z)u_{s}(w) &\sim& \frac{-1}{z-w} u_{s}'(w) + {\it terms~in~other~
powers~of~}\frac{1}{z-w}, \nonumber\\
u_{1}(z)u_{s}(w) &\sim& \frac{-2}{z-w} [\sum^{s}_{l=1}(-1)^{l}
\left( \begin{array}{c}
s\\l
\end{array} \right) u_{0}^{(l)}u_{s-l} +\frac{u_{s+1}'}{(s+1)}
+\frac{(-1)^{s}u_{0}^{(s+2)}}{(s+1)(s+2)}](w) \nonumber\\
& & +{\it terms~in~other~powers~of~}\frac{1}{z-w}
\end{eqnarray}
where $u^{(l)}\equiv \partial^{l} u$. Note that
the terms in other powers of $(z-w)^{-1}$ do not to contribute to the
commutators between charges after double integration. Eq.(17) are essential to
our following construction of commuting $\hat{W}_{\infty}$ charges.

Before proceeding, let us recall
some definitions for the local (normal-ordered) product of several
local operators. First the local product $(AB)(z)$ of two local operators
$A(z)$ and $B(z)$ is defined by
\begin{eqnarray}
(AB)(z) = \oint_{z} \frac{A(w)B(z)}{w-z} dw
\end{eqnarray}
in which the small contour of integration encircles $z$. This local product
is non-commutative, i.e. $(AB)(z)\neq (BA)(z)$; however they differ from
each other only by total derivatives:
\begin{eqnarray}
\oint_{0} (AB-BA)(z) dz =0.
\end{eqnarray}
According to the definition (18), the operator product of $C(z)$
with the local product $(AB)(w)$ is given by
\begin{eqnarray}
C(z)(AB)(w) = ((C(z)A(w))B(w)) + (A(w)(C(z)B(w))).
\end{eqnarray}
The multiple local product of several operators are generally non-associative,
e.g.
\begin{eqnarray}
(A(BC))-(B(AC)) = ((AB)C)-((BA)C).
\end{eqnarray}
Finally we define the symmetric local product of $N$ local
operators to be the totally symmetrized sum of their multiple local products
{\it taken from the left}:
\begin{eqnarray}
\langle A_{1}A_{2}\cdots A_{N} \rangle = \frac{1}{N!}\sum_{P\{i\}}
(\cdots ((A_{i_{1}}A_{i_{2}})A_{i_{3}})\cdots A_{i_{N}})
\end{eqnarray}
where $P\{i\}$ denotes the summation over all possible permutations.

Now we turn to construct an infinite set of mutually commuting
quantum $\hat{W}_{\infty}$ charges (with $p=1$).
Let us assign a degree 1 to $\partial_z$ and $r+2$ to $u_r$ and
assume that the $m$-th quantum charge density $H_{m}(z)$
is homogeneous and of degree $m$ and is led by a term linear in the
highest-spin current $u_{m-2}$ with unit coefficient. Therefore the most
general form for $H_m(z)$ is
\begin{eqnarray}
H_{m}(z) = \sum_{l}\sum_{\{i,a\}}C_{i_{1}i_{2}\cdots i_{l}}^{a_{1}a_{2}
\cdots a_{l}}(m) (\cdots (u_{i_{1}}^{(a_{1})}u_{i_{2}}^{(a_{2})})\cdots
u_{i_{l}}^{(a_{l})}) (z), ~~~~m=2,3,\ldots
\end{eqnarray}
where $l=1,2,\cdots,[m/2]$ is the number of currents $u_{r}$ in the
product (the maximal value of $l$ being the integral part of $m/2$);
$\{i,a\}$ stands for the set of all possible indices satisfying
$i_{1}+i_{2}+\cdots +i_{l}+a_{1}+a_{2}+\cdots +a_{l}$ $=m-2l$ and
$C_{i_{1}i_{2}\cdots i_{l}}^{a_{1}a_{2}\cdots a_{l}}(m)$ are constant
coefficients. We want to determine these coefficients
so that the corresponding charges $Q_{m}\equiv \oint H_{m}(z)dz$
commute with each other:
\begin{eqnarray}
{[\oint_{0} H_{n}(z)dz , \oint_{0} H_{m}(w)dw]} = 0.
\end{eqnarray}
Especially, for the $n=2$ case, the current $H_2$, by definition,
is nothing but the energy-momentum tensor $u_0$. It is easy to see
from the first equation of (17) and eq.(20) that
the commutativity with $Q_2$, i.e.
\begin{eqnarray}
{[\oint_{0} H_{2}(z)dz , \oint_{0} H_{m}(w)dw]} = 0,
\end{eqnarray}
is always satisfied. The first set of nontrivial equations in eq.(24)
start with $n=3$:
\begin{eqnarray}
{[\oint_{0} H_{3}(z)dz , \oint_{0} H_{m}(w)dw]} = 0.
\end{eqnarray}
Here $H_{3}$ can only be $u_{1}$ plus a derivative of $u_{0}$;
the latter does not contribute to the charge $Q_3$. In the following
we will show that all the charge density $H_{m}(z)$ modulo total
derivatives, and thus all charges $Q_{m}$, are completely determined
by eq.(26) alone. Both amusingly and amazingly the so-determined
charges $Q_{m}$ automatically commute with each other; in other words,
eqs.(24) are automatically satisfied by the solution to eq.(26).

First let us consider how to solve eq.(26).  A straightforward calculation,
making extensive use of the second equation of (17),
determines {\it uniquely} the first seven charges from eq.(26), giving
\begin{eqnarray}
& & Q_{2} = \oint u_{0}(z)dz, ~~~~Q_{3} = \oint u_{1}(z)dz, \nonumber\\
& & Q_{4} = \oint (u_{2}-u_{0}u_{0})(z)dz, ~Q_{5} =
\oint (u_{3}-6u_{0}u_{1})(z)dz, \nonumber\\
& & Q_{6} = \oint (u_{4}-12u_{0}u_{2}-12u_{1}u_{1}+8(u_{0}u_{0})u_{0})(z)dz,
\nonumber\\
& & Q_{7} = \oint (u_{5}-20u_{0}u_{3}-60u_{1}u_{2}+60(u_{0}u_{0})u_{1}
+60(u_{0}u_{1})u_{0})(z)dz, \nonumber\\
& & Q_{8} = \oint (u_{6}-30u_{0}u_{4}-120u_{1}u_{3}-90u_{2}u_{2}
+180(u_{0}u_{0})u_{2} +180(u_{0}u_{2})u_{0} \nonumber\\
& & ~~~~~~~~+360(u_{1}u_{0})u_{1}+360(u_{1}u_{1})u_{0}
-180((u_{0}u_{0})u_{0})u_{0})(z)dz.
\end{eqnarray}
In principle, the explicit construction of charges that satisfy eq.(26)
may successively continue to higher orders with increasing labor and
effort. However, by inspecting eq.(27), it is amusing to observe the
following simple and nice features:
(a) No term containing any derivative of the
currents $u_{r}$ appears in these charges at all.
(b) We note this is true only when all local products of
currents are chosen to start from the left; if they had started
from the right, terms with derivatives of $u_{r}$ would
appear. (We do not feel fully
understand this phenomenon.) (c) For every charge in eq.(27), the
coefficients in the right-hand side conspire to result in
totally symmetric multiple products of currents involved, after applying
eqs.(19) and (21). Generalizing these empirical rules, we are led to
the ans\'atz that the infinite set of
charges we are looking for to solve eq.(26) are of the simple form
\begin{eqnarray}
Q_{m} = \oint_{0} \sum_{l} \sum_{\{i\}}C_{i_{1}i_{2}\cdots i_{l}}(m)
\langle u_{i_{1}}u_{i_{2}}\cdots u_{i_{l}} \rangle (z)dz
\end{eqnarray}
where $l=1,2,\cdots, [m/2]$ is the number of currents in the symmetrized
product, which we call the level of the term. (Do not confuse it
with the level of the model.) For given $l$, the summation is over
all partitions $\{i_{k}\}$ satisfying $i_{1}+i_{2}+\cdots
+i_{l}=m-2l$ and $i_{1}=i_{2}=\cdots =i_{d_{1}}$ $<i_{d_{1}+1}=\cdots
=i_{d_{1}+d_{2}}$ $<\cdots =i_{d_{1}+d_{2}+\cdots +d_{k}(=l)}$; here
$d$'s denote the degeneracies in the $i$'s.
We have been able to find an elegant formula for the
coefficients that nicely summarizes our lower-order results (27):
\begin{eqnarray}
C_{i_{1}i_{2}\cdots i_{l}}(m) = \frac{(-1)^{l-1}(l-1)!(m-2)!}{d_{1}!d_{2}!
\cdots d_{k}!i_{1}!i_{2}!\cdots i_{l}!}.
\end{eqnarray}
The coefficient of the leading linear term $u_{m-2}$
is unity as desired. One feels the success of this formula
should not be accidental. Indeed eq.(29) provides a
unique solution to eq.(26) for arbitrary $m$.
Here we give a sketchy description of the proof.

Using eq.(17) and performing one integration, we rewrite eq.(26) as
\begin{eqnarray}
& & \oint_{0} \sum_l \sum_{\{i\}}C_{i_{1}i_{2}\cdots i_{l}}(m)
\sum^{l}_{a=1}
\langle u_{i_{1}}\cdots u_{i_{a-1}} [\sum^{i_{a}}_{k=1}(-1)^{k}
\left( \begin{array}{c}
i_{a}\\k
\end{array} \right) (u_{0}^{(k)}u_{i_{a}-k}) \nonumber\\
& & +\frac{u_{i_{a}+1}'}{(i_{a}+1)}
+\frac{(-1)^{i_{a}}u_{0}^{(i_{a}+2)}}{(i_{a}+1)(i_{a}+2)}] u_{i_{a+1}}
\cdots u_{i_{l}} \rangle (z)dz = 0.
\end{eqnarray}
Note that each term has only one derivative
on one of the currents. To verify eq.(29), one
needs to show that all the terms in eq.(30) with the same
order of derivative on one of the $u_{r}$'s must cancel each other. In
particular, collecting the terms with a first-order derivative on one of the
currents we want to verify that
\begin{eqnarray}
\oint_{0} \sum_l \sum_{\{i\}}C_{i_{1}i_{2}\cdots i_{l}}(m)\sum^{l}_{a=1}
\langle u_{i_{1}}\cdots u_{i_{a-1}} {[-i_{a}(u_{0}'u_{i_{a}-1})
+\frac{u_{i_{a}+1}'}{(i_{a}+1)}]}
u_{i_{a+1}}\cdots u_{i_{l}} \rangle (z)dz = 0.
\end{eqnarray}
At first sight how to handle the local product $(u_{0}'u_{i_{a}-1})$ in the
first term in the middle of a symmetrized product seems to be a big problem.
Fortunately, by applying repeatedly the relations (19) and (21),
we have been able to prove the following lemma:
\begin{eqnarray}
\oint_{0} \langle (A_{0}A_{1})\cdots A_{i}\cdots A_{N} \rangle =
\oint_{0} \langle A_{0}A_{1}\cdots A_{i}\cdots A_{N} \rangle
\end{eqnarray}
where the left-hand side is symmetrized with respect to the $N$ indices $(1,2,
\cdots,N)$ and the right-hand side the $N+1$ indices $(0,1,2,\cdots,N)$.
Using this to rearrange the factors in eq.(31) and balancing the
number of factors, we have for each $l$:
\begin{eqnarray}
& & \oint_{0} \sum_{\{i\}_{l+1}}C_{i_{1}^{d_{1}}i_{2}^{d_{2}}
\cdots i_{k}^{d_{k}}}(m)\sum^{k}_{a=1} \frac{d_{a}}{i_{a}+1}
u_{i_{1}}^{d_{1}}\cdots u_{i_{a}}^{d_{a}-1}
u_{i_{a}+1}' u_{i_{a+1}}^{d_{a+1}} \cdots u_{i_{k}}^{d_{k}} (z)dz
\nonumber\\
&=& \oint_{0} \sum_{\{j\}_{l}}C_{j_{1}^{c_{1}}j_{2}^{c_{2}}
\cdots j_{n}^{c_{n}}}(m)\sum^{n}_{a=1} c_{a}j_{a}u_{0}'
u_{j_{1}}^{c_{1}}\cdots u_{j_{a}-1}u_{j_{a}}^{c_{a}-1}
u_{j_{a+1}}^{c_{a+1}} \cdots u_{j_{n}}^{c_{n}} (z)dz
\end{eqnarray}
where the $C$'s on the left-hand side are level-$(l+1)$ coefficients and
those on the right-hand side level-$l$ ones: $\sum d_{a} =l+1$,
$\sum d_{a} i_{a}=m-2(l+1)$, and $\sum c_{b} =l$, $\sum c_{a} j_{b}= m-2l$,
with $i_{1}<i_{2}<\cdots <i_{k}$, $j_{1}<j_{2}<\cdots <j_{n}$.

We are going to prove the validity of this equation by induction for both
the level $l$ and the first index $i_1$.
For terms on the left-hand side having derivative on the
highest-spin current $u_{i_{k}}$, we do integration by parts and turn such
terms into those containing no derivative on $u_{i_{k}}$.
On the left-hand side there appears a term
with the first index $i_1=0$ and of the form
$u_{0}'u_{0}^{d_{1}-1}u_{i_{2}}^{d_{2}}\cdots u_{i_{k-1}}^{d_{k-1}}
u_{i_{k}}^{d_{k}-1}u_{i_{k}+1}$, with coefficient
\begin{eqnarray}
-\frac{d_{1}d_{k}}{i_{k}+1}C_{0^{d_{1}}i_{2}^{d_{2}}
\cdots i_{k}^{d_{k}}}(m).
\end{eqnarray}
Terms of the same form on the right-hand side have the coefficients
\begin{eqnarray}
& & C_{0^{d_{1}-2}1i_{2}^{d_{2}}\cdots i_{k-1}^{d_{k-1}}i_{k}^{d_{k}-1}
(i_{k}+1)}(m) +(i_{2}+1) C_{0^{d_{1}-1}i_{2}^{d_{2}-1}(i_{2}+1)
i_{3}^{d_{3}}
\cdots i_{k-1}^{d_{k-1}}i_{k}^{d_{k}-1}(i_{k}+1)}(m) \nonumber\\
& & +\cdots +(i_{k-1}+1) C_{0^{d_{1}-1}i_{2}^{d_{2}}
\cdots i_{k-1}^{d_{k-1}-1}(i_{k-1}+1)i_{k}^{d_{k}-1}(i_{k}+1)}(m)
\nonumber\\
& & +2(i_{k}+1) C_{0^{d_{1}-1}i_{2}^{d_{2}}
\cdots i_{k-1}^{d_{k-1}}i_{k}^{d_{k}-2}(i_{k}+1)^{2}}(m)
+(i_{k}+2) C_{0^{d_{1}-1}i_{2}^{d_{2}}
\cdots i_{k-1}^{d_{k-1}}i_{k}^{d_{k}-1}(i_{k}+2)}(m) \nonumber\\
&=& \frac{(-1)^{l-1}l!(m-2)!}{(d_{1}-1)!d_{2}!\cdots d_{k-1}!(d_{k}-1)!
(i_{2}!)^{d_{2}}\cdots (i_{k-1}!)^{d_{k-1}}(i_{k}!)^{d_{k}-1}(i_{k}+1)!}.
\end{eqnarray}
Here we have assumed the validity of eq.(29) for level $l$. The
equality between (34) and (35) requires exactly the validity of eq.(29)
with $i_1=0$ at level $l+1$. Furthermore, we need to verify that after
the above-mentioned integration by parts,
all the terms having derivative
on $u_{i}$ with $i\neq 0$ on the left-hand side of eq.(33) cancel
each other. For example, consider the terms of the form
$u_{i_{1}}'u_{i_{1}}^{d_{1}-1}u_{i_{2}}^{d_{2}}\cdots u_{i_{k-1}}^{d_{k-1}}
u_{i_{k}}^{d_{k}-1}u_{i_{k}+1}$. Their coefficients are
\begin{eqnarray}
-\frac{d_{1}d_{k}}{i_{k}+1}C_{i_{1}^{d_{1}}i_{2}^{d_{2}}\cdots
i_{k}^{d_{k}}}(m) +\frac{1}{i_{1}}C_{(i_{1}-1)i_{1}^{d_{1}-1}i_{2}^{d_{2}}
\cdots i_{k-1}^{d_{k-1}}i_{k}^{d_{k}-1}(i_{k}+1)}(m).
\end{eqnarray}
Assuming eq.(29) is true for the level-$(l+1)$ coefficient with the first
index $i_{1}-1$, then the vanishing of (36) yields the correct
level-$(l+1)$ coefficient $C_{i_{1}^{d_{1}}i_{2}^{d_{2}}\cdots
i_{k}^{d_{k}}}(m)$ with the first index $i_{1}(\neq 0)$.
Similarly we have checked the cancellation of
all other terms, particularly those
having derivative on higher-spin currents $u_{i_{a}}$ (with $i_{a}
\neq 0, i_{k}$)
on the left-hand side of eq.(33).
Thus, the validity of eq.(31) or (33) is verified with the coefficient
$C$'s given by eq.(29).

With eq.(31) verified, eq.(30) reduces to
\begin{eqnarray}
& & \oint_{0}\sum_l \sum_{\{i\}}C_{i_{1}i_{2}\cdots i_{l}}(m)\sum^{l}_{a=1}
\langle u_{i_{1}}\cdots u_{i_{a-1}} [\sum_{k=2}^{i_{a}} (-1)^{k}
\left( \begin{array}{c}
i_{a}\\k
\end{array} \right) (u_{0}^{(k)}u_{i_{a}-k}) \nonumber\\
& & +\frac{(-1)^{i_{a}}u_{0}^{(i_{a}+2)}}{(i_{a}+1)(i_{a}+2)}] u_{i_{a+1}}
\cdots u_{i_{l}} \rangle (z)dz = 0.
\end{eqnarray}
Again we need to separate out terms having the same order $k$ for the
derivative on one of the currents and show the cancellation among them.
This can be done in a way very similar to above discussion, and
we leave such details to a longer publication [15]. It is not hard to
convert the above verification into an inductive determination of the
expression (29) for the coefficient $C$'s in the ans\'atz (28),
starting from the normalized coefficient at level $l=1$. If we had set the
coefficient (of the leading linear term $u_{m-2}$) to be zero, then
all other coefficients in eq.(30) should vanish.

Thus we have proved that eq.(29) uniquely solves eq.(30) or eq.(26) under
the ans\'atz (28). Furthermore, we have been able to prove
that this solution is actually the only solution to the commutativity
equation (26) under the very general assumption (23).
In fact let us add to this solution a
term $\oint_{0} p_{m}(z)dz$, with $p_{m}(z)$ arbitrary homogeneous
polynomial of $u_{r}$ and their derivatives of total degree $m$, and require
\begin{eqnarray}
{[Q_{3},\oint_{0}p_{m}(z)dz ]} =0.
\end{eqnarray}
(Obviously, $p_{3}=0$ up to total derivatives.)
By an inductive argument similar to what we have before, the only solution
containing at least one derivative of the currents is
\begin{eqnarray}
\oint_{0}p_{m}(z)dz = 0.
\end{eqnarray}

Now we need to prove that the commutativity (26)
of $Q_m$'s with $Q_3$ will guarantee their mutual commutativity (24).
This might look a bit too strong, but a similar situation
happened for the infinite set of commuting charges for quantum KdV equation
[16]. In fact this is a consequence of the Jacobi identities
\begin{eqnarray}
{[Q_{3}, {[Q_{m}, Q_{n}]}]} + {[Q_{m}, {[Q_{n}, Q_{3}]}]} +
{[Q_{n}, {[Q_{3}, Q_{m}]}]} = 0.
\end{eqnarray}
It follows immediately from eq.(26) that
\begin{eqnarray}
{[Q_{3}, {[Q_{m}, Q_{n}]}]} = 0.
\end{eqnarray}
Here we note that both $Q_{m}$ and $Q_{n}$ are homogeneous and of degree
$m-1$ and $n-1$ respectively. Besides the OPE's in the $\hat{W}_{\infty}$
algebra
are homogeneous, so is the commutator $[Q_{m}, Q_{n}]$ with degree $m+n-2$.
Thus, $[Q_{m}, Q_{n}]$ must be an integral of something which is of
the general form (23). The above-proved uniqueness of the homogeneous
solution (28) plus (29) assures us that
in view of eq.(41), the commutator $[Q_{m}, Q_{n}]$
must be proportional to $Q_{m+n-1}$ up to a constant factor:
\begin{eqnarray}
{[Q_{m}, Q_{n}]} = c~ Q_{m+n-1}.
\end{eqnarray}
We note that on the left-hand
side the charge density $H_{m+n-1}$ is led by the linear term $u_{m+n-3}$
and does not involve any term containing derivatives of currents.
On the other hand, as a general feature of the $\hat{W}_{\infty}$
algebra (14)-(15),
the commutator between densities $H_{m}$ and $H_{n}$, led by $u_{m-2}$ and
$u_{n-2}$ respectively, does not give rise to
the desired $u_{m+n-3}$ term or any term with no derivatives on
currents.\footnote[1]{This is obvious from the classical $\hat{W}_{\infty}$
algebra (5-6) and it appears to survive quantization.}
Therefore, the constant $c$ in eq.(42) must be zero, yielding eq.(24).
(As a check, we have explicitly verified that the charges $Q_{m}$
in eq.(27) are really commuting with each other up to $m=5$.)

We emphasize that in the above proof starting from eq.(23),
besides the OPE's between the $\hat{W}_{\infty}$ currents $u_{r}$,
nowhere we have used the two-boson realization (14) or (16)
of these currents.

In the present $SL(2,R)/U(1)$ model at
level $k$=1, the existence of an infinite set of commuting quantum
charges $Q_{m}$ gives rise to a huge infinite-dimensional symmetry
in this model. In fact with the help of these charges we can
generate an infinite dimensional compatible flow of the basic
bosonic currents $j(z)$ and $\bar{j}(z)$ as follows:
\begin{eqnarray}
\frac{\partial j}{\partial t_{m}} =
{[j, Q_{m}]}, ~~~~
\frac{\partial \bar{j}}{\partial t_{m}} =
{[\bar{j}, Q_{m}]}.
\end{eqnarray}
It is easy to see that the fundamental OPE's between $j$ and
$\bar{j}$
\begin{eqnarray}
\bar{j}(z)j(w) &\sim& \frac{1}{(z-w)^{2}}, \nonumber\\
\bar{j}(z)\bar{j}(w) &\sim& j(z)j(w) ~\sim ~ 0
\end{eqnarray}
are invariant under these flows. For example, for the OPE $\bar{j}j$, we have
(at $t_{m}=0$)
\begin{eqnarray}
\frac{\partial (\bar{j}j)}{\partial t_{m}} = {[\bar{j}, Q_{m}]}j
+\bar{j}{[j, Q_{m}]} ={[\bar{j}j, Q_{m}]} \sim {[\frac{1}{(z-w)^{2}}, Q_{m}]}
=0.
\end{eqnarray}
Moreover, the flows of $j$ and $ \bar{j}$ induce similar flows of the
composite $\hat{W}_{\infty}$ currents
\begin{eqnarray}
\frac{\partial u_{r}}{\partial t_{m}} =
{[u_{r}, Q_{m}]},
\end{eqnarray}
under which the $\hat{W}_{\infty}$ current algebra for
$u_{r}$ is invariant as well. We point out that the commuting charges
$Q_{m}=\oint H_{m}(z)dz$ are constant along the flows they generate:
\begin{eqnarray}
\frac{\partial Q_{n}}{\partial t_{m}} = {[Q_{n}, Q_{m+1}]} = 0.
\end{eqnarray}
Thus the flows (46) they generate are compatible:
\begin{eqnarray}
\frac{\partial^{2} u_{r}}{\partial t_{m}\partial_{n}} =
\frac{\partial^{2} u_{r}}{\partial t_{n}\partial_{m}}.
\end{eqnarray}
And eq.(47) implies the integrability of the flows (46).

We observe that the charges we construct in this paper are simple integrals,
$\oint_{0} H_m(z)dz$, of the Hamiltonian currents $H_{m}$.
Though our spin-2 current $H_{2}(z)$ is indeed
the energy-momentum tensor of the model, however our charge $Q_{2}$,
given by eq.(27), is the charge $L_{-1}$ in usual
notation, rather than the usual energy (or the Hamiltonian) $L_{0}$.
Therefore what we have obtained is
an infinite set of mutually commuting quantum charges containing
$L_{-1}$. Though they are not
conserved charges in the model (since they do not commute with $L_{0}$),
they are still interesting charges in that they
generate a huge symmetry maintaining the basic current algebra
for $j$ and $\bar{j}$ and the composite $\hat{W}_{\infty}$
current algebra invariant. We speculate that when we go from the conformal
model to string theory, these symmetries would turn into an infinite
set of compatible $\hat{W}_{\infty}$ constraints.

To conclude, three more remarks are in order. First, we expect
the existence of an infinite set of commuting quantum $\hat{W}_{\infty}$
charges which include the Hamiltonian $L_{0}$ of the model. Secondly,
it should be possible to construct a Hamiltonian (out of two bosons) for
which the set of charges we have constructed in this letter become genuine
conserved charges; and we expect such Hamiltonian may represent a perturbed
conformal but integrable field theory. We hope to address these problems
in future publications. Finally, the similarity of eq.(46) with the
classical KP flows (9) suggests that the former can be viewed as a
quantum version or quantum deformation of the latter, so that the flows
(46) are eligible to be called as quantum KP flows. The justification
needs to go beyond the $k=1$ case, and is left to another publication
of ours [15].

\vspace{5 pt}
{\it{Acknowledgements}}:~ The authors thank C. J. Zhu for
discussions. The work is supported in part by NSF grant PHY-9008452.

\vspace{40 pt}
\begin{center}
{\large REFERENCES}
\end{center}
\begin{itemize}
\vspace{5 pt}

\item[1.] A. A. Belavin, A. M. Polyakov and A. B. Zamolodchikov, Nucl. Phys.
B241 (1984) 333.
\item[2.] V. Knizhnik and A. B. Zamolodchikov, Nucl. Phys. B247 (1984) 83;
D. Gepner and E. Witten, Nucl. Phys. B278 (1986) 493.
\item[3.] A. B. Zamolodchikov, Theor. Math. Phys. 65 (1985) 1205.
\item[4.] Vl. S. Dotsenko and V. A. Fateev, Nucl. Phys. B240 [FS 12]
(1984) 312; B. L. Feigin and D. B. Fuchs, Func. Aual. Appl. 16 (1982);
ref. [1].
\item[5.] E. Witten, Comm. Math. Phys. 92 (1986) 455.
\item[6.] P. Goddard, A. Kent and D. Olive, Phys. Lett. B152 (1985) 88;
Comm. Math. Phys. 103 (1986) 105.
\item[7.] F. Yu and Y.-S. Wu, Nucl. Phys. B373 (1992) 713.
\item[8.] F. Yu and Y.-S. Wu, Phys. Rev. Lett. 68 (1992) 2996.
\item[9.] I. Bakas and E. Kiritsis, Maryland/Berkeley/LBL preprint
UCB-PTH-91/44, LBL-31213 and UMD-PP-92/37, Sept.1991.
\item[10.] I. Bakas, Phys. Lett. B228 (1989) 57; Comm. Math. Phys. 134
(1990) 487.
\item[11.] C. Pope, L. Romans and X. Shen, Phys. Lett. B236 (1990) 173;
B242 (1990) 401; F. Yu and Y.-S. Wu, Phys. Lett. B263 (1991) 220;
K. Yamagishi, Phys. Lett. B259 (1991) 436.
\item[12.] M. Sato, RIMS Kokyuroku 439 (1981) 30; E. Date, M. Jimbo, M.
Kashiwara and T. Miwa, in Proc. of RIMS Symposium on Nonlinear Integrable
Systems, eds. M. Jimbo and T. Miwa, (World Scientific, Singapore, 1983);
G. Segal and G. Wilson, Publ. IHES 61 (1985) 1.
\item[13.] L. A. Dickey, Annals New York Academy of Sciences, (1987) 131.
\item[14.] A. Gerasimov, A. Marshakov and A. Morozov, Nucl. Phys. B328
(1989) 664.
\item[15.] F. Yu and Y.-S. Wu, in preparation.
\item[16.] R. Sasaki and I. Yamanaka, Commun. Math. Phys. 108 (1987) 691;
Adv. Stud. in Pure Math. 16 (1988) 271; B. Feigin and E. Frenkel,
RIMS/Harvard preprint RIMS-827, Oct.1991; P. Di Francesco, P. Mathieu and
D. S\'en\'echal, Princeton/Laval preprint PUPT-1300/ LAVAL-PHY-28/91,
Dec.1991.

\end{itemize}

\end{document}